\documentstyle[12pt]{article}
\begin{document}
\baselineskip 20pt        
\noindent
\hspace*{13cm}
hep-ph/9811379\\
\noindent
\hspace*{13cm}
KIAS-P98034\\
\noindent
\hspace*{13cm}
YUMS 98-018\\  
\noindent
\hspace*{13cm}
SNUTP 98-125\\ 
\vspace{1.8cm}

\begin{center}
{\Large \bf  Bi-maximal Lepton Flavor Mixing Matrix \\
and Neutrino Oscillation}\\

\vspace*{1cm}

{\bf Sin Kyu Kang $^{a,}$\footnote{skkang@kias.re.kr, skkang@supy.kaist.ac.kr}},~~
{\bf C. S. Kim $^{a,b,}$\footnote{kim@cskim.yonsei.ac.kr,~~
http://phya.yonsei.ac.kr/\~{}cskim/}},

\vspace*{0.5cm}

 $a:$ School of Physics, Korea Institute for Advanced Study, Seoul 130-012, Korea \\
 $b:$ Physics Department, Yonsei University, Seoul 120-749, Korea \\ 
        
\vspace{1.0cm}
 
(\today)
\vspace{0.5cm}

\end{center}        

\begin{abstract}
\vspace{0.2cm}

\noindent
Recently, many authors showed that if the solar and atmospheric neutrino 
data are both described by maximal mixing vacuum oscillations at the relevant 
mass scale, then there exists a unique bi-maximal lepton mixing matrix 
for three neutrino flavors.
We construct the lepton mass
matrices from the symmetry principle so that maximal mixings
for the atmospheric and the solar neutrino vacuum 
oscillations are naturally generated.
Although the hierarchical patterns of the lepton sector are quite different from
each other, we show how two different mass matrices suggested in this work
can be generated in a unified way.
We also give comments on possible future tests of the bi-maximal lepton mixing
matrix.

\end{abstract}

\vspace{0.3cm}
PACS number(s): 14.60Pq, 12.15Fk

\newpage
The recent atmospheric neutrino data  from the Super-Kamiokande 
Collaboration \cite{superk1} presents convincing evidence for 
neutrino oscillation and hence nonzero neutrino mass.
The results indicate the maximal mixing between $\nu_{\mu}$ and
$\nu_{\tau}$ with mass squared difference $\delta m^2_{atm}\simeq
5\times 10^{-3}~ \mbox{eV}^2$.
The long-standing solar neutrino deficit  \cite{solar1,solar2,solar3}
can also be explained  through the 
matter enhanced neutrino oscillation  ({\it i.e.} the MSW solution 
\cite{msw}) if $\delta m^2_{solar}\simeq 6\times 10^{-6}~\mbox{eV}^2$ and
$\sin^2 2\theta_{solar}\simeq 7\times 10^{-3}$ (small angle case), or
$\delta m^2_{solar}\simeq 9\times 10^{-6}~\mbox{eV}^2$ and
$\sin^2 2\theta_{solar}\simeq 0.6$ (large angle case)
and through the long-distance vacuum neutrino oscillation called
as ``just-so" oscillation \cite{justso} if $\delta m^2_{solar}\simeq 10^{-10}
~\mbox{eV}^2$ and $\sin^2 2\theta_{solar}\simeq 1.0$.
However, the recent data on the electron neutrino spectrum reported by 
Super-Kamiokande \cite{solar2} seems to favor the "just-so" vacuum oscillation, 
even though the small angle MSW oscillation and the maximal mixing between the
atmospheric $\nu_{\mu}$ and $\nu_{\tau}$ have been taken as a natural solution
for the neutrino problems \cite{yana}.
Moreover, as shown by Georgi and Glashow \cite{georgi}, solar neutrino oscillations
may be nearly maximal if relic neutrinos comprise at least one percent
of the critical mass density of the universe.
If this vacuum oscillation of the solar neutrino is confirmed in future 
experiments \cite{solar2,sno}, 
the mixing angles in the lepton sector are turned out to be large
in contrast with the quark sector in which all observed mixing angles
among different families are quite small.
This seems not to be achieved in such a way to unify quarks and leptons
at the GUT scale.
One can thus deduce that the origin of the lepton mass matrices would be
different from the one of the quark sector \cite{yana}.
Therefore, it is worthwhile to find any possible mechanism providing
such  neutrino mixing patterns.  
Gauge models such as $SO(10)$ grand unification model \cite{yana2} and 
left-right symmetric model  \cite{moha}
have been constructed so that the so called ``bi-maximal" neutrino mixing 
\cite{bimax}
for the solar and atmospheric vacuum oscillations are naturally accommodated.
There have also been attempts to derive such a neutrino mixing from
a lepton mass matrix ansatz \cite{georgi,bimax,bimax2,xing}.

Recently, Barger {\it et al.} \cite{bimax} showed that 
if the solar and atmospheric neutrino 
data are both described by maximal mixing vacuum oscillations at the relevant 
mass scale, then there exists a unique mixing matrix for three neutrino flavors.
Their solution necessarily conserves CP and automatically implies that there 
is no disappearance of atmospheric $\nu_e$, consistent with indications from 
the Super-Kamiokande experiment. However, they did not construct the neutrino 
mass matrix
from some simple symmetry principle, but inverted the process to
obtain the neutrino mass matrix in the flavor basis from the mass eigenvalues
and the bi-maximal mixing matrix by using the fact that
a Majorana mass matrix or a hermitian Dirac mass matrix 
can be diagonalized by a single unitary matrix.
{}From the phenomenological point of view, Georgi and Glashow \cite{georgi}
also suggested 
the neutrino mass matrix that is compatible with the ``bi-maximal" neutrino
mixing, cosmological observation and the nonexistence of neutrinoless double
beta decay.

The purpose of this letter is to construct the lepton mass
matrices from the symmetry principle so that maximal mixings 
for the atmospheric and the solar neutrino vacuum 
oscillations are naturally generated.
We note that the bi-maximal lepton flavor mixing matrix $V_{bi-max}$ 
can be constructed from
the product of two unitary matrices
\begin{eqnarray}
\left(U^{lepton}_{CKM}\right)^{\dagger}\equiv V_{bi-max} &\equiv& \left( \begin{array}{ccc} 
\frac{1}{\sqrt{2}} & -\frac{1}{\sqrt{2}} & 0 \\
\frac{1}{2} & \frac{1}{2} & -\frac{1}{\sqrt{2}} \\
\frac{1}{2} & \frac{1}{2} &   \frac{1}{\sqrt{2}}
\end{array} \right) \\
 &=& \left( \begin{array}{ccc}
1 & 0 & 0 \\
0 & \frac{1}{\sqrt{2}} & -\frac{1}{\sqrt{2}} \\
0 & \frac{1}{\sqrt{2}} & \frac{1}{\sqrt{2}} \end{array} \right) \cdot
\left( \begin{array}{ccc}
\frac{1}{\sqrt{2}} & -\frac{1}{\sqrt{2}} & 0 \\
 \frac{1}{\sqrt{2}} & \frac{1}{\sqrt{2}} & 0 \\
0 & 0 & 1 \end{array} \right) 
\equiv U_\nu^{\dagger} \cdot U_l ,
\end{eqnarray}
where $U_\nu$ and $U_l$ give the maximal mixing
between the second and the third generations
and between the first and the second generations, respectively.
As will be shown later,
the charged lepton mass matrix can be diagonalized by $U_l$, while
the neutrino mass matrix can be diagonalized by $U_\nu$.
This is outstanding feature of our lepton mass matrices.
Although the hierarchical patterns of the lepton sector are quite different from
each other, we will show how two different mass matrices suggested in this work
can be generated in a unified way.

Let us start with a general $S(3)_L\times S(3)_R$ symmetric mass matrix 
\cite{kang,tani}:
\begin{eqnarray}
M_0=C\left( \begin{array}{ccc}
      1 & r & r \\
      r & 1 & r \\
      r & r & 1 \end{array} \right) .
\end{eqnarray}
By diagonalizing this matrix with the help of the unitary matrix
\begin{eqnarray}
U=\left( \begin{array}{ccc}
      \frac{1}{\sqrt{2}} & -\frac{1}{\sqrt{2}} & 0 \\
      \frac{1}{\sqrt{6}} & \frac{1}{\sqrt{6}} & -\frac{2}{\sqrt{6}} \\
      \frac{1}{\sqrt{3}} & \frac{1}{\sqrt{3}} & \frac{1}{\sqrt{3}} 
       \end{array} \right),
\end{eqnarray}
we obtain the eigenvalues 
$$
C(1-r,~~ 1-r,~~ 1+2r).
$$
For $r=1$, only the third element becomes massive, which enables us to explain
why the third generation quarks and charged leptons are much heavier than 
the others \cite{fritz}. Thus we take $r=1$ for the charged lepton mass matrix.
On the other hand, the neutrino data does not seem to support such hierarchy.
Moreover, if we regard the neutrinos as a part of hot dark matter,
all three neutrinos may be almost degenerate in their masses \cite{mass,csk}.
This almost degenerate neutrino mass pattern can be achieved by taking
$r$ to be nearly zero \cite{tani}.
Therefore, we choose, as the first step
\footnote{Actually, the case $r=0$ might 
not  require the diagonalization of $M_0$ because that case corresponds to the 
diagonal form already before diagonalizing. However, we need the 
diagonalization as long as the value of $r$ is not exactly zero but small 
enough  to be negligible compared to the parameter $C$ and even $A,B$.},
\begin{eqnarray}
r &=& 1~~~{\rm for~ charged~ lepton~ case}, \nonumber\\
{\rm and}~~~r &=& 0~~~{\rm for~ neutrino~ case}. \nonumber
\end{eqnarray}
In order to generate the hierarchy of the charged lepton sector and 
phenomenologically acceptable form of the mass matrix for the neutrino sector,
we introduce the symmetry breaking terms so that the
hierarchy or the mass difference between two generations
can be accommodated, and the maximal mixing between those generations
can be generated simultaneously.
We will show that this can be achieved in the way that the
$S(3)_L \times S(3)_R$ symmetry is broken down to $S(2)_L \times S(2)_R$.

Now we consider 
the following $2\times 2$ mass matrix, which provides 
the maximal mixing between two flavors \cite{twobi}
\begin{eqnarray}
\left( \begin{array}{cc}
 \alpha & \beta  \\
 \beta & \alpha  
 \end{array} \right ) .
\end{eqnarray}
This form of mass matrix can be diagonalized by the unitary matrix
\begin{eqnarray}
U=\frac{1}{\sqrt{2}}\left( \begin{array}{cc}
 1 & -1  \\
 1 & 1  
 \end{array} \right ) ,
\end{eqnarray}
and the eigenvalues are given as 
$$
(\alpha+\beta,~~ \alpha-\beta) .
$$
The matrix (5) can be easily generated by  
considering the so-called ``democratic" $2\times 2$ 
mass matrix, that reflects $S(2)_L \times S(2)_R$ symmetry, and
by adding a symmetry breaking matrix, 
which has $S(2)$
symmetry under the interchange between the first and the second indices:
\begin{eqnarray}
 M_2 =A\left( \begin{array}{cc}
      1 & 1 \\
      1 & 1 \end{array} \right)
  + B\left( \begin{array}{cc}
      1 & -1 \\
      -1 & 1 \end{array} \right) 
     =\left( \begin{array}{cc}
      A+B & A-B \\
      A-B & A+B \end{array} \right) .
\end{eqnarray}
With the help of Eq. (6), one can easily obtain the eigenvalues of $M$ which
are given as 
$$
(2A,~~ 2B) .
$$

Since we want to get the bi-maximal mixing matrix 
while keeping the hierarchical charged lepton masses and degenerate neutrino masses,
we add this symmetry breaking matrix $M_2$ to the previous 
hierarchical matrices $M_0$ appropriately. And then,
relate the parameters $A$ and $B$ 
\begin{itemize}
\item (a) to the masses of
the first and the second generations for the charged lepton sector, 
respectively, and
\item (b) to the mass differences between two (the second and the third) 
generations for the neutrino sector.
\end{itemize}
At the end,  we can obtain the realistic lepton mass matrices.
{\it I.e.}, we add the above symmetry breaking $M_2$
matrix as the sub matrix of $M_0$ in the ($e, \mu$) basis 
for the charged lepton sector; 
\begin{eqnarray}
 M_l= \left(\begin{array} {ccc}
      0 & 0 & 0\\
      0 & 0 & 0\\
      0  & 0  & C \end{array} \right) \Rightarrow
 \left( \begin{array}{ccc}
      A+B & A-B & 0\\
      A-B & A+B & 0\\
      0  & 0  & C \end{array} \right),
\end{eqnarray}
while as the one in the ($\nu_{\mu}, \nu_{\tau}$) basis 
for the neutrino sector as follows;
\begin{eqnarray}
  M_{\nu}= \left(\begin{array} {ccc}
       C & 0 & 0\\
       0 & C & 0\\
       0  & 0  & C \end{array} \right) \Rightarrow
  \left( \begin{array}{ccc}
      C & 0 & 0 \\
      0 & C+A+B & A-B \\
      0 & A-B & C+A+B \end{array} \right) .
\end{eqnarray}
Then, one can see that these matrices $M_l$ and $M_\nu$ can be 
diagonalized by $U_l$ and
$U_\nu$, respectively, which in turn lead to the bi-maximal lepton flavor
mixing matrix $U_{CKM}^{lepton}$, 
as given in Eqs.(1,2) by combining $U_\nu$ with $U_l$.

Eigenvalues of the mass matrices $M_l$ and $M_{\nu}$ are given as 
$$
M_l = (2A,~~ 2B,~~ C)~~~ {\rm and} ~~~M_\nu = (C,~~ C+2A,~~ C+2B) ,
$$ 
respectively.
For the charged lepton sector, the parameters $A,B$ and $C$ are determined by
the following mass relations
\begin{equation}
A=m_e/2, ~~~B=m_{\mu}/2~~~ {\rm and} ~~~C=m_{\tau} .
\end{equation}
In order to solve  $A,B$ and $C$ for the neutrino sector,  we first require
two conditions,
$$
\Delta m^2_{solar}=10^{-10}~\mbox{eV}^2,~~~{\rm and}~~~ \Delta m^2_{atm}=2\times
10^{-3}~\mbox{eV}^2 ,
$$
which can fit the available data quite well, where 
the mass differences
$\Delta m^2_{ij}=m^2_{\nu_i}-m^2_{\nu_j}$ should be identified with, 
among the possibilities,
$\Delta m^2_{solar}=\Delta m^2_{12}$ and $\Delta m^2_{atm}=\Delta m^2_{23}$.
Thus we will consider henceforth only this case. 
In addition, if the neutrinos account for the hot dark matter of the universe,
one has to require 
$$
\sum|m_{\nu_i}|\simeq 6~ \mbox{eV}.
$$
Then the set of parameters $(A,B,C)$ is given by 
\begin{equation}
(A,B,C)\approx (10^{-10},~~ 0.00025,~~ 1.9999)~ (\mbox{eV})~ ,
\end{equation}
for which three light neutrinos are almost degenerate with masses around
$2$ eV.

Now, we check if the solution of three neutrino mass eigenvalues satisfies
the constraints coming from the neutrinoless double $\beta-$decay,
as well as other data from neutrino oscillation experiments.   
The neutrino mixing matrix Eq.~(1) and neutrino mass eigenvalues 
lead to
$$
\langle m_{\nu_e} \rangle \equiv  \left| \sum_{i=1}^3 V_{ei}^2 m_i 
\right|^2 \simeq 1.9999 ~{\rm eV} .
$$
However, that value of neutrino mass is not compatible with the current upper 
limit coming from the non-observation of the
neutrinoless double $\beta-$decay, which is given as \cite{bb}
\begin{equation}
\langle m_{\nu_e}\rangle \leq (0.5 - 1.5) ~~\mbox{eV}.
\end{equation}
In order to be satisfied with this constraint, $\sum |m_{\nu_i}|$ is
allowed only up to $4.5$ eV.
If we take this value, the set of parameters $(A,B,C)$ is determined
to be
\begin{equation}
(A,B,C)\approx (10^{-10},~~ 0.00035,~~ 1.4998)~ (\mbox{eV})~,
\end{equation}
for which three light neutrinos are almost degenerate with masses around
$1.5$ eV.
If we begin to increase the neutrino masses in order 
to make them dominant hot dark matter candidates, we cease to satisfy the 
$(\beta\beta)_{\nu 0}$ constraint.

Further test of our ansatz is provided with the long baseline experiments
searching for $\nu_{\mu} \rightarrow \nu_{\tau}$ oscillation in the range 
of  $\Delta m_{\mu\tau}^2 \simeq 10^{-3}~{\rm eV}^2$ \cite{kkkk}. 
The MINOS \cite{minos} and K2K \cite{k2k} sensitivities to $\Delta m^2$
at $90\%$ CL can go down to $\Delta m^2 = 1.2 \times 
10^{-3}~\mbox{eV}^2$ and $2.0\times 10^{-3}~\mbox{eV}^2$, respectively,
while the ICARUS \cite{ica} sensitivity is achieved at $\Delta m^2=3.0\times 10^{-3}~
\mbox{eV}^2$.
The bi-maximal mixing scenario, in which $\sin^2 2\theta_{\mu \tau}$ is predicted
to be $1$ with $\Delta m^2_{\mu \tau}\simeq 2\times 10^{-3}~\mbox{eV}^2$,
can be tested at the MINOS and K2K experiments searching for the
$\nu_{\mu} \rightarrow \nu_{\tau}$ oscillations in the foreseeable future,
but is beyond the sensitivity to $\Delta m^2$ at $90\%$ CL being achieved 
at ICARUS.
Future experiment on the $\nu_{\mu} \leftrightarrow \nu_{\tau}$ oscillation
from the MINOS and K2K will exclude our model for charged lepton and neutrino
mass matrices.

Finally, we comment on that the bi-maximal neutrino mixing matrix Eq. (1) 
predicts zero for $V_{e3}$ element
which makes $\nu_{e} \leftrightarrow \nu_{\mu}$ and 
$\nu_{\mu} \leftrightarrow \nu_{\tau}$ oscillations to be effectively
a two-channel problem.  This is supported from CHOOZ data \cite{chooz}
which give the mixing angle $\theta_{13}$ to be less than $13^{0}$ in
most of the Super-Kamiokande allowed region.
As one can see, $V_{e3}$ element becomes zero in the limit of $\theta_{13}=0$
\cite{pecci}.
However, note that a non-vanishing $V_{e3}$ element is not completely excluded,
but rather it can be larger in the region not covered by 
CHOOZ \cite{chooz2,lisi}.
To justify this bi-maximal mixing scenario, the precise determination
of $V_{e3}$ element will may be essential, which requires several oscillation
channels to be probed at the same time.
{}From the fact that the $\nu_{\mu} \rightarrow
\nu_{\tau}$ disappearance channel is sensitive only to $V_{\mu3}^2$ and
the $\nu_{\mu} \rightarrow \nu_{e}$ appearance channel is sensitive to
the product $V_{\mu 3}^2 V_{e3}^2$,  one can determine the element $V_{e3}$
by combining the regions to be probed in both channels.
K2K \cite{k2k} will be expected to perform this, but it does not, at present,
seem to achieve sufficient
sensitivity in the $\nu_{\mu} \rightarrow \nu_{e}$ appearance channel to
probe the region of $V_{e3}^2$ allowed by Super-Kamiokande and CHOOZ \cite{lisi}.
\\

{\Large \bf Acknowledgments}
\\

\noindent
CSK wishes to thank the Korea Institute for Advanced Study for warm
hospitality.
CSK wishes to acknowledge the financial
support of Korean Research Foundation made in the program of 1997.

\newpage
%\vspace{0.5cm}


\begin{thebibliography}{99}

\bibitem{superk1} Super-Kamiokande Collaboration, Talk by T. kajita at {\it 
Neutrino-98}, Takayama, Japan, June 1998.

\bibitem{solar1} B. T. Cleveland {\it et al.}, Nucl. Phys. B(Proc. Suppl.) 
{\bf 38}, 47 (1995); 
Kamiokande Collaboration, Y. Fukuda {\it et al.}, 
Phys. Rev. Lett. {\bf 77}, 1683 (1996); 
GALLEX Collab., W. Hampel {\it et al.}, Phys. Lett. {\bf B 388}, 384 (1996); 
SAGE Collaboration, J. N. Abdurashitov {\it et al.}, 
Phys. Rev. Lett. {\bf 77}, 4708 (1996).

\bibitem{solar2} SuperKamiokande Collaboration, 
Talk by Y. Suzuki at {\it Neutrino-98}, 
Takayama, Japan, June 1998.

\bibitem{solar3} 
J. N. Bahcall and M. H. Pinsonneault, Rev. Mod. Phys. {\bf 67}, 781 (1995); 
J. N. Bahcall, S. Basu and M. H. Pinsonneault, astro-ph/9805135.

\bibitem{msw} L. Wolfenstein, Phys. Rev. {\bf D 17}, 2369 (1978); 
S. P. Mikheyev and A. Smirnov, Yad. Fiz. {\bf 42}, 1441 (1985); 
Nuovo Cimento {\bf 9 C}, 17 (1986).

\bibitem{justso} 
V. Barger, R. J. N. Phillips and K. Whisnant, Phys. Rev. {\bf D24}, 538 (1981);
S. L. Glashow and L. M. Krauss, Phys. Lett. {\bf B190}, 199 (1987).

\bibitem{yana} T. Yanagida, Talk at {\it Neutrino-98}, Japan, June 1998;
 P. Ramond, Talk at {\it Neutrino-98}, Japan, June 1998.

\bibitem{georgi} H. Georgi and S. L. Glashow, hep-ph/9808293.

\bibitem{sno} SNO Collaboration, Talk by A. McDonald at  {\it Neutrino-98}, Japan, June 1998.

\bibitem{yana2} Y. Nomura and T. Yanagida, hep-ph/9807325.

\bibitem{moha} R. N. Mohapatra and S. Nussinov, hep-ph/9808301.

\bibitem{bimax} V. Barger, S. Pakvasa, T. J. Weiler and K. Whisnant, 
hep-ph/9806387.

\bibitem{bimax2}  R. N. Mohapatra and  S. Nussinov, hep-ph/9809415.

\bibitem{xing} H. Fritzsch and Z. Z. Xing, Phys. Lett. {\bf B 440}, 313 (1998),
   hep-ph/9808272.

\bibitem{kang} K. Kang, S. K. Kang, J. E. Kim and P. Ko, , Mod. Phys. Lett. A
               {\bf 12}, 553 (1997), hep-ph/9611369; K. Kang and S. K. Kang,
               hep-ph/9802328.

\bibitem{tani} H. Fritzsch and Z. Z. Xing, Phys. Lett. {\bf B 372}, 265 (1996);
              hep-ph/9807234 (talk given at the Ringberg Euroconference on New
              Trends in Neutrino Physics, Ringberg, Germany, May 1998);
              M. Tanimoto, hep-ph/9807515;
              M. Fukugita, M. Tanomoto and T. Yanagida, Phys. Rev. {\bf D57},
            4429 (1998).

\bibitem{fritz} A. C. Rothman and K. Kang, Phys. Rev. Lett. {\bf 43}, 1548 (1979); 
H. Fritzsch, Phys. Lett. {\bf B 73}, 317 (1978); Nucl. Phys. {\bf B 155}, 189
(1979); K. Kang and S. K. Kang, Phys. Rev. {\bf D56}, 1511 (1997); 
hep-ph/9802330 and references therein.

\bibitem{mass} D. O. Caldwell and R. N. Mohapatra, 
Phy. Rev. {\bf D 48}, 3259 (1993); S. T. Petcov and A. Smirnov, 
Phys. Lett. {\bf B 322}, 109 (1994); A. S. Joshipura, 
Z. fur Phys. {\bf C 64}, 31 (1994); A. Ionissian and J. W. F. Valle, 
Phys. Lett. {\bf B 332}, 93 (1994); P. Bamert and C. P. Burgess, 
Phys. Lett. {\bf B 329}, 289 (1994); D. G. Lee and R. N. Mohapatra, 
Phys. Lett. {\bf B 329}, 463 (1994); R. N. Mohapatra and S. Nussinov, 
Phys. Lett. {\bf B 346}, 75 (1995);
P. Harrison, D. Perkins and W. Scott, Phys. Lett. {\bf B 349}, 137 (1995); 
A. Acker and S. Pakvasa, Phys. Lett. {\bf B 397}, 209 (1997); 
P. Krastev and S. Petcov, Phys. Lett. {\bf B 395}, 69 (1997); 
J. Peltoniemi and J. W. F. Valle, Nucl. Phys. {\bf B 406}; 
S. M. Bilenky, C. Giunti and W. Grimus, hep-ph/9607372; hep-ph/9711311; 
C. Cardall and G. Fuller, astro-ph/9606024; 
C. Cardall, D. Cline and G. Fuller, hep-ph/9706426.

\bibitem{csk}
K. Kang, S. K. Kang, C. S. Kim and S. M. Kim, hep-ph/9808419.
 
\bibitem{twobi} K. Kang, S. K. Kang, J. E. Kim and P. Ko, Phys. Lett. {\bf B}
in press, hep-ph/9706535;
   see also R. Barbieri, L. J. Hall, D. Smith, A. Smith, A. Strumia and
   N.Weiner, hep-ph/9807235; 
   R. Barbieri, L. J. Hall and A. Strumia; hep-ph/9808333.

\bibitem{bb}Heidelberg-Moscow Collab., M. Gunther {\it et al.}, 
Phys. Rev. {\bf D 55}
(1997) 54. Note that the uncertainty in the bound of Eq. (12) reflects an
uncertainty in the calculation of the relevant nuclear matrix element.

\bibitem{kkkk} See also K. Kang, S. K. Kang, J. E. Kim and P. Ko,  
hep-ph/9706535.

\bibitem{minos} Minos collaboration, {\it P-875:  A long baseline neutrino
               oscillation experiment at Fermilab}, NuMI-L-63, Feb. 1995.

\bibitem{k2k} M. Sakuda, K2K collaboration, {\it The KEK-PS Long Baseline
              Neutrino Oscillation Experiment(E362)},  talk given at the
               workshop Pacific Particle Physics Phenomenology,
             Seoul National University,
      Seoul, Korea, 31 October-2 November, 1997.

\bibitem{ica} A. Rubbia, Nucl. Phys. B (Proc. Suppl.) {\bf 66}, 436 (1998).
\bibitem{chooz} CHOOZ Collaboration, M. Apollonio {\it et al.}, hep-ex/9711002.
\bibitem{pecci}R. D. Peccei, Summary talk at the XXIX International Conference
on High Energy Physics, Vancouver, Canada July 1998; see also X. -Y. Pham,
hep-ph/9809322.

\bibitem{chooz2} 
M. C. Gonzalez-Garcia, H. Nunokawa, O. L. G. Peres, J. W. F. Valle, hep-ph/9807305; 
S. M. Bilenkii, C. Giunti and W. Grimus, hep-ph/9809368;
R. Barbieri, L. J. Hall, D. Smith, A. Strumia and N. Weiner, hep-ph/9807235;
G. Altarelli and F. Feruglio, hep-ph/9809596.
\bibitem{lisi} G. L. Fogli, E. Lisi, A. Marrone and G. Scioscia, hep-ph/9808205.
\end{thebibliography}
\end{document}